\documentclass[12pt]{article}

\usepackage{amsmath,amsfonts,amssymb}
\usepackage{epsfig}
\begin{document}

\begin{titlepage}

\begin{center}

\vskip 0.4 cm

\begin{center}
{\Large{ \bf Hamiltonian Formalism of Bimetric Gravity In Vierbein
Formulation  }}
\end{center}

\vskip 1cm

{\large Josef Kluso\v{n}$^{}$\footnote{E-mail: {\tt
klu@physics.muni.cz}} }

\vskip 0.8cm

{\it Department of
Theoretical Physics and Astrophysics\\
Faculty of Science, Masaryk University\\
Kotl\'{a}\v{r}sk\'{a} 2, 611 37, Brno\\
Czech Republic\\
[10mm]}

\vskip 0.8cm

\end{center}

\begin{abstract}
This paper is devoted to the Hamiltonian analysis of
 bimetric gravity in vierbein formulation. We identify all constraints and
 determine their nature. We also show an existence of additional
 constraint so that the scalar mode can be eliminated.

\end{abstract}

\bigskip

\end{titlepage}

\newpage


\def\tSigma{\tilde{\Sigma}}
\def\mM{\mathcal{M}}
\def\tepsilon{\tilde{\epsilon}}
\def\mR{\mathcal{R}}
\def\bare{\bar{e}}
\def\tmW{\tilde{\mathcal{W}}}
\newcommand{\bC}{\mathbf{C}}
\newcommand{\bD}{\mathbf{D}}
\def\hf{\hat{f}}
\def\tK{\tilde{K}}
\def\tmK{\tilde{\mK}}
\def\mS{\mathcal{S}}
\def\tn{\tilde{n}}
\def\mP{\mathcal{P}}
\def\bA{\mathbf{A}}
\def\hE{\hat{E}}
\newcommand{\eI}[3]{e_{(#1)_#2}^{\ \quad #3}}
\def\bF{\mathbf{F}}
\def\bV{\mathbf{V}}
\def\bW{\mathbf{W}}
\def\hF{\hat{F}}
\def\hE{\hat{E}}
\def\bN{\bar{N}}
\def\mT{\mathcal{T}}
\def\bP{\bar{P}}
\def\bJ{\mathbf{J}}
\def\bpi{\bar{\pi}}
\def\tlambda{\tilde{\lambda}}
\def\bB{\mathbf{B}}
\def\bnabla{\bar{\nabla}}
\def\bN{\bar{N}}
\def\bE{\mathbf{E}}
\def\bGamma{\bar{\Gamma}}
\def\baK{\bar{K}}
\def\pp{{\, \mid \hskip -1.5mm =}}
\def\bg{\bar{g}}
\def\bL{\mathbf{L}}
\def\cL{\mathcal{L}}
\def\tnabla{\tilde{\nabla}}
\def\tx{\tilde{x}}
\def\tmG{\tilde{\mG}}
\def\mC{\mathcal{C}}
\def\be{\begin{equation}}
\def\bD{\mathbf{D}}
\def\ee{\end{equation}}
\def\bD{\mathbf{D}}
\def\bea{\begin{eqnarray}}

\def\eea{\end{eqnarray}}
\def\bk{\mathbf{k}}
\def\tr{\mathrm{tr}\, }
\def\tmH{\tilde{\mH}}
\def\tY{\mathcal{Y}}
\def\nn{\nonumber \\}
\def\bI{\mathbf{I}}
\def\tmV{\tilde{\mV}}
\def\e{\mathrm{e}}
\def\bE{\mathbf{E}}
\def\bX{\mathbf{X}}
\def\bY{\mathbf{Y}}
\def\bR{\bar{R}}
\def\hN{\hat{N}}
\def\hK{\hat{K}}
\def\hnabla{\hat{\nabla}}
\def\hc{\hat{c}}
\def\mH{\mathcal{H}}
\def \Gi{\left(G^{-1}\right)}
\def\hZ{\hat{Z}}
\def\bz{\mathbf{z}}
\def\bK{\mathbf{K}}
\def\iD{\left(D^{-1}\right)}
\def\tmJ{\tilde{\mathcal{J}}}
\def\tr{\mathrm{Tr}}
\def\mJ{\mathcal{J}}
\def\partt{\partial_t}
\def\parts{\partial_\sigma}
\def\bG{\mathbf{G}}
\def\str{\mathrm{Str}}
\def\Pf{\mathrm{Pf}}
\def\bM{\mathbf{M}}
\def\tA{\tilde{A}}
\newcommand{\mW}{\mathcal{W}}
\def\bx{\mathbf{x}}
\def\by{\mathbf{y}}
\def \mD{\mathcal{D}}
\newcommand{\tZ}{\tilde{Z}}
\newcommand{\tW}{\tilde{W}}
\newcommand{\tmD}{\tilde{\mathcal{D}}}
\newcommand{\tN}{\tilde{N}}
\newcommand{\hC}{\hat{C}}
\newcommand{\hg}{\hat{g}}
\newcommand{\hX}{\hat{X}}
\newcommand{\bQ}{\mathbf{Q}}
\newcommand{\hd}{\hat{d}}
\newcommand{\tX}{\tilde{X}}
\newcommand{\calg}{\mathcal{G}}
\newcommand{\calgi}{\left(\calg^{-1}\right)}
\newcommand{\hsigma}{\hat{\sigma}}
\newcommand{\hx}{\hat{x}}
\newcommand{\tchi}{\tilde{\chi}}
\newcommand{\mA}{\mathcal{A}}
\newcommand{\ha}{\hat{a}}
\newcommand{\tB}{\tilde{B}}
\newcommand{\hrho}{\hat{\rho}}
\newcommand{\hh}{\hat{h}}
\newcommand{\homega}{\hat{\omega}}
\newcommand{\mK}{\mathcal{K}}
\newcommand{\hmK}{\hat{\mK}}
\newcommand{\hA}{\hat{A}}
\newcommand{\mF}{\mathcal{F}}
\newcommand{\hmF}{\hat{\mF}}
\newcommand{\hQ}{\hat{Q}}
\newcommand{\mU}{\mathcal{U}}
\newcommand{\hPhi}{\hat{\Phi}}
\newcommand{\hPi}{\hat{\Pi}}
\newcommand{\hD}{\hat{D}}
\newcommand{\hb}{\hat{b}}
\def\I{\mathbf{I}}
\def\tW{\tilde{W}}
\newcommand{\tD}{\tilde{D}}
\newcommand{\mG}{\mathcal{G}}
\def\IT{\I_{\Phi,\Phi',T}}
\def \cit{\IT^{\dag}}
\newcommand{\hk}{\hat{k}}
\def \cdt{\overline{\tilde{D}T}}
\def \dt{\tilde{D}T}
\def\bra #1{\left<#1\right|}
\def\ket #1{\left|#1\right>}
\def\mV{\mathcal{V}}
\def\Xn #1{X^{(#1)}}
\newcommand{\Xni}[2] {X^{(#1)#2}}
\newcommand{\bAn}[1] {\mathbf{A}^{(#1)}}
\def \bAi{\left(\mathbf{A}^{-1}\right)}
\newcommand{\bAni}[1]
{\left(\mathbf{A}_{(#1)}^{-1}\right)}
\def \bA{\mathbf{A}}
\def\ta{\tilde{a}}
\newcommand{\Qn}[1]{Q^{(#1)}}
\newcommand{\Fn}[1]{F^{(#1)}}
\newcommand{\An}[1]{A^{(#1)}}
\def\tpsi{\tilde{\psi}}
\newcommand{\bT}{\mathbf{T}}
\def\bmR{\bar{\mR}}
\newcommand{\ba}{\mathbf{a}}
\newcommand{\tphi}{\tilde{\phi}}
\newcommand{\mL}{\mathcal{L}}
\newcommand{\mbQ}{\mathbf{Q}}
\def\mat{\tilde{\mathbf{a}}}
\def\bai{(\mathbf{a}^{-1})}
\def\mati{(\tilde{\mathbf{a}}^{-1})}
\def\mtF{\tilde{\mathcal{F}}}
\def \tZ{\tilde{Z}}
\def\mtC{\tilde{C}}
\def\hmD{\hat{\mD}}
\def\hnabla{\hat{\nabla}}
\def \tY{\tilde{Y}}
\def\pb #1{\left\{#1\right\}}
\newcommand{\E}[3]{E_{(#1)#2}^{ \quad #3}}
\newcommand{\p}[1]{p_{(#1)}}
\newcommand{\hEn}[3]{\hat{E}_{(#1)#2}^{ \quad #3}}
\def\mbPhi{\mathbf{\Phi}}

\section{Introduction}
Bimetric theories of gravity are based on the idea to join the two
tensors $\hg_{\mu\nu}$ and $\hf_{\mu\nu}$ in a symmetric way when
each tensor has its own Einstein-Hilbert action and then couple
these actions through a non-derivative mass term. The presence of
this term reduces the separate coordinate invariances to a single
one \cite{Salam:1969rq,Isham:1970gz} \footnote{Bimetric theories of
gravity were studied intensively in the past, see for example
\cite{Chamseddine:1978yu,Chamseddine:2004dh,Chamseddine:2003ft}.}.
If we set one metric as the background metric without any dynamics
we find that the bimetric theory is reduced to a single metric
massive gravity theory with the mass term that at the linear limit
leads to the Fierz-Pauli free theory \cite{Fierz:1939ix}.
 However it was shown soon
that this theory propagates ghost modes at non-linear level
\cite{Boulware:1973my,Boulware:1972zf}. On the other hand new form
of the massive term was proposed recently in
\cite{deRham:2010kj,deRham:2011rn,Hassan:2011hr,Hassan:2011vm} that
was shown to be ghost free even at the non-linear level
\cite{Hassan:2011ea,Hassan:2012qv}, see also
\cite{Kluson:2012wf,Comelli:2013txa,Comelli:2013paa}.

This form of the massive gravity was further generalized in
\cite{Hassan:2011tf} where the dynamical gravity was coupled to the
general reference metric. Then it was the small step to the
generalization of given construction to the bimetric gravity when
the fixed reference metric becomes dynamical with its own
Einstein-Hilbert action \cite{Hassan:2011zd}. It was also argued
there and in \cite{Hassan:2011ea} that this theory is ghost free.
However this analysis was discussed  in \cite{Kluson:2013cy} where
it was argued that the analysis performed in \cite{Hassan:2011ea}
does not show an existence of the additional constraint in case of
the bimetric gravity \footnote{The Hamiltonian analysis of bimetric
gravity was also performed in
\cite{Kluson:2013lza,Kluson:2012ps,Soloviev:2012wr,Soloviev:2013mia}.}.

The non-linear massive gravity and bimetric gravity that are claimed
that are ghost free are based on the specific form of the potential
that contains the square root of
$\hg^{\mu\nu}\hf_{\nu\rho}$. This is rather awkward structure
 which makes very
difficult to find an extra constraint that could eliminate the
Boulware-Deser ghost. However as was shown in beautiful paper
\cite{Hinterbichler:2012cn} the square root structure suggests that
the vierbein variables $E_\mu^{ \ A}$ could be the appropriate one
for the formulation of consistent bimetric theories. In more
details, completely new multimetric interacting spin-2 theories were
proposed in \cite{Hinterbichler:2012cn} using the powerful vierbein
formulation of the general relativity and corresponding mass terms.
It was argued there that due to the specific form of the interaction
terms the action is linear in lapses and shifts which implies an
existence of additional constraints that could eliminate
non-physical modes. However  we mean that the Hamiltonian analysis
presented in given paper  was not complete. In particular, the
constraints corresponding to the diagonal diffeomorphism were not
identified and it was not shown that they are the first class
constraints.

The goal of this paper is to fill this gap and perform the
Hamiltonian analysis of the bimetric gravity in vierbein formulation
with the simplest form of the potential between two vierbeins
$E^\mu_{ \ A}$ and $F_\mu^{ \ A}$. We explicitly show that it is
crucial to analyze the time developments of the constraints
corresponding broken spatial rotation. It is also important to
stress that when we use the parametrization of the vierbein as in
\cite{Hinterbichler:2012cn} we should interpret $p_a$-that will be
defined below-as a dynamical variable with no time dependence in the
action. As a result the conjugate momentum vanishes and is the
primary constraint of the theory. Then the requirement of the
preservation of given constraint leads to another secondary
constraint that was not consider in the literature so far.

Very important  point is to identify the constraints that are
generators of diagonal diffeomorphism. To do this we follow
\cite{Damour:2002ws} when we introduce new variables that are
functions  of $N,N^i$ and $M,M^i$ which are lapses and shifts in
$\hg_{\mu\nu}$ and $\hf_{\mu\nu}$ respectively.
 Then we determine eight new secondary constraints
where four of them should correspond to the generators of diagonal
diffeomorphism on condition that  the Poisson brackets between new
Hamiltonian constraint $\bmR$ closes on the constraints surface. The
similar analysis was performed previously in case of bimetric
gravity formulated with metric variables in
\cite{Kluson:2013lza,Kluson:2012ps}. It turns out that in case of
  bimetric theory in vierbein formulation the situation is more
complicated and we have to take into account the presence of the new
secondary constraints. Then we are able to show that the Poisson
bracket between Hamiltonian constraints vanish on the constraint
surface. On the other hand one can ask the question why we use the
variables introduced in  \cite{Damour:2002ws} in case of bimetric
gravity formulated with metric variables in case of the bimetric
gravity formulated using vierbein formalism. The answer is that we
mean that they are the only variable where it is possible to
identify generators of diagonal diffeomorphism that is difficult to
identify with the help of another choose of variables. Further,
using this formalism we can easily see an analogy between bimetric
theory formulated using either metric of vierbein variables.

With the help of this result we proceed to the analysis of the
consistency of all constraints during the time development of the
system. Now due to the very remarkable structure of the vierbein
formulation of bi-gravity we find an existence of additional
constraint which leads to the elimination of the scalar mode in the
same way as in case of non-linear massive gravity
\cite{Hassan:2012qv,Hassan:2011ea}. This result confirms the results
derived in \cite{Alexandrov:2013rxa}. More precisely, in
\cite{Alexandrov:2013rxa} canonical analysis of bimetric gravity
formulated in vierbein formalism  where the spin connection is
treated as an independent field was performed with elegant
formulations of the secondary constraints that are responsible for
the elimination of the ghosts. On the contrary our analysis is more
closely related to the formulation introduced in
\cite{Hinterbichler:2012cn} where the spin connection is not
considered as an independent field however the constraints
responsible for the elimination of ghost are much more complicated.

This paper is organized as follows. In the next section
(\ref{second}) we introduce the bimetric gravity in vierbein
formalism and find its Hamiltonian, identify all constraints and
determine their constraint structure. In Conclusion (\ref{third}) we
outline our results. Finally in Appendix we review the Hamiltonian
formulation of general relativity action formulated in the vierbein
formalism.
\section{Vierbein Formulation of Bimetric Gravity}\label{second}
General vierbein can by written in the upper triangular form
  and we denote this vierbein with hat
\begin{equation}\label{utf}
\hE_\mu^{ \ A}=\left(\begin{array}{cc} N & N^ie_i^{ \ a} \\
0 & e_i^{ \ a} \\ \end{array}\right) \ , \quad \hE^\mu_{ \ A}=
\left(\begin{array}{cc} \frac{1}{N} & 0 \\
-\frac{N^i}{N} & e^i_{ \ a} \\ \end{array}\right)
\end{equation}
where $N$ and $N^i$ are the $4$ time-like components. The spatial
vielbeins $e_i^{ \ a}$ contain $9$ components that are related to
the spatial part of the metric by
\begin{equation}
g_{ij}=e_i^{ \ a}e_j^{ \ b}\delta_{ab} \ .
\end{equation}
Now by writing out the metric of this vierbein we find
\begin{eqnarray}
\hg_{\mu\nu}=E_\mu^{ \ A}E_\nu^{ \ B}\eta_{AB}=
\left(\begin{array}{cc} -N^2+N^iN_i & N_i \\
N_j & g_{ij} \\ \end{array}\right) \ , \quad \eta_{AB}=
\mathrm{diag}(-1,1,1,1)
\end{eqnarray}
that means that $N$ and $N^i$ are the usual lapse and shifts that
appear in the ADM decomposition of the metric
\cite{Gourgoulhon:2007ue,Arnowitt:1962hi,Chaichian:2011sx}. Note
that the inverse metric has the form
\begin{eqnarray}
\hg^{\mu\nu}=\hE^\mu_{ \ A}\hE^\nu_{ \ B}\eta^{AB} \ .  \nonumber \\
\end{eqnarray}
Then by definition \footnote{For review of vierbein formalism, see
\cite{Yepez:2011bw}.}
\begin{eqnarray}
\hE_\mu^{ \ A}\hE^\nu_{ \ A}
&=& \delta_\mu^{ \ \nu} \ , \quad  \hE_\mu^{ \ A}\hE^\mu_{ \ B}=
\delta^A_{ \ B} \ , \nonumber \\
   e_i^{ \ a}e^j_{ \ a}&=& \delta_i^{
\ j} \ , \quad e_i^{ \ a}e^i_{ \ b}=\delta^a_{ \ b} \ .
\nonumber \\
\end{eqnarray}
The upper triangular form does not fix the local Lorentz invariance
since it leaves a residual spatial rotation. There are $4$
components in $N,N^i$ and $9$ in the spatial vielbein. The remaining
$3$ components of the vielbein have been fixed by using the upper
triangular gauge choice. It is possible to formulate an arbitrary
vierbein as the action of same boost on the upper triangular
vierbein. Note that for $3-$dimensional  vector $p_a$ we define a
standard Lorentz boost as
\begin{equation}
\Lambda(p)^A_{ \ B}=\left(\begin{array}{cc} \gamma & p_b \\
p^a & \delta^a_{ \ b}+\frac{1}{\gamma+1}p^a p_b \\ \end{array}
\right) \ , \gamma=\sqrt{1+p_ap^a} \ ,
\end{equation}
where $p^a=\delta^{ab}p_b$ and  where by definition
\begin{equation}
\eta_{AB}\Lambda^A_{ \ C}\Lambda^B_{ \ D}=\eta_{CD}
\end{equation}
so that
\begin{eqnarray}
\left(\Lambda^{-1}\right)^A_{ \ B}= \left(\begin{array}{cc} \gamma & -p^b \\
-p_a & \delta_a^{ \ b}+\frac{p_a p^b}{\gamma+1} \\
\end{array}\right) \ .
\end{eqnarray}
The boost takes the $4-$dimensional vector $(1,0,0,0)$ into the unit
normalized $4-$vector
\begin{equation}
\Lambda^A_{  \ B}\left(\begin{array}{cc} 1 \\
0 \\ \end{array}\right)=
\left(\begin{array}{cc} \gamma \\
p^a \\ \end{array}\right) \ .
\end{equation}
Then we write the general vierbein as the standard boost of the
upper triangular vierbein
\begin{eqnarray}\label{genvier}
E_\mu^{  \ A}=\Lambda(p)^A_{ \ B}\hE_\mu^{ \ B}=
\left(\begin{array}{cc} N+N^ie_i^{ \ a}p_a & Np^b+ N^ie_i^{
\ a}(\delta_a^{ \ b}+ \frac{1}{\gamma+1}p_a p^b) \\
e_i^{  \ a}p_a & e_i^{ \ a}(\delta_a^{ \ b}+ \frac{1}{\gamma+1}p_a
p^b)\end{array}\right) \ .
 \nonumber \\
\end{eqnarray}
  We see that $16$ components of the
general vierbein are now parameterized by the $4$ components of $N$
and $N^i$ together with $9$ components of the spatial vielbein
$e_i^{ \ a}$ and $3$ components of $p_a$.

It is important that the  Einstein-Hilbert action is invariant under
local Lorentz transformation. As a result it is possible to
partially fix the gauge and express the Einstein-Hilbert action
using the upper triangular form. This fact greatly simplifies the
Hamiltonian formalism of general relativity in vierbein formalism.
 The detailed analysis is
performed in the Appendix \ref{appendix}.

Now we are ready to proceed to the vierbein formulation of the
bimetric gravity   when we
 consider bigravity with two metrics
\begin{equation}
\hg_{\mu\nu}=E_\mu^{ \ A}E_\nu^{ \ B}\eta_{AB} \ , \quad
\hf_{\mu\nu}=F_\mu^{ \ A}F_\nu^{ \ B}\eta_{AB} \
\end{equation}
with Einstein-Hilbert actions for both of these metrics. Then
without the interaction term the action is invariant under two
separate local Lorentz transformations
\begin{equation}
E'^{ \ A}_\mu(x)= \Lambda_{(g) B}^{ \ A}(x)E_{\mu}^{ \ B}(x) \ ,
\quad  F'^{\ A}_{\mu}(x)= \Lambda_{(f) B}^{ \ A}(x)F_{\mu}^{\ B}(x)
\ .
\end{equation}
The action is also invariant under two diffeomorphisms
\begin{equation}
E'^{ \ A}_{\mu}(x')df_{(1)}^\mu =E_{\nu}^{ \ A}(x)dx^\nu \ , F'^{ \
A}_{\mu}(x')df_{(2)}^\mu=F_{\nu}^{ \ A}(x)dx^\nu \ .
\end{equation}
Then we consider the action in the form \cite{Hinterbichler:2012cn}
\begin{eqnarray}
S&=&\frac{M_g^{2}}{2} \int d^4x (\det E)R[E] + \frac{M_f^{2}}{2}
\int d^4x (\det F) R[F] -
\nonumber \\
&-&\mu^2 \int d^4x \sum_{n=0}^4 \beta_n (\det E)
S_n(E^{-1}F) \ ,  \nonumber \\
\end{eqnarray}
where $\mu^2=\frac{1}{8}m^2 M_{fg}^{2}$ and
 where $S_n$ are symmetric polynomials whose explicit
definitions can be found in \cite{Hinterbichler:2012cn}. It was
shown here that they can  be written in terms of traces of the matrix
$\mathbb{M}$ as
\begin{eqnarray}
S_0(\mathbb{M})&=&1 \ , \nonumber \\
S_1(\mathbb{M})&=&[\mathbb{M}] \nonumber \\
S_2(\mathbb{M})&=&\frac{1}{2!}([\mathbb{M}]^2-[\mathbb{M}^2])
 \ ,
 \nonumber \\
 S_3(\mathbb{M})&=&\frac{1}{3!}
 ([\mathbb{M}]^3-3[\mathbb{M}][\mathbb{M}^2]+2[\mathbb{M}^3]) \ ,
 \nonumber \\
 S_4(\mathbb{M})&=&
 \frac{1}{4!}
 ([\mathbb{M}]^4-6[\mathbb{M}^2][\mathbb{M}]^2+8[\mathbb{M}][\mathbb{M}^3]
 +3[\mathbb{M}^2]^2-6[\mathbb{M}^4]) \ ,
 \nonumber \\
 \end{eqnarray}
 where $[\mathbb{M}]$ means the trace of the matrix $\mathbb{M}$. In what follows we
 restrict ourselves to the simplest non-trivial case corresponding
 to $\beta_0=\beta_2=\beta_3=\beta_4=0\ , \beta_1=1$ which however
 captures the main property of given theory.

Now we proceed to the Hamiltonian analysis of the bimetric theory in
the vierbein formulation. We use the parametrization of the general
vierbein introduced in (\ref{genvier}). Explicitly
\begin{eqnarray}
E_{\mu}^{ \ A}=\Lambda(p)^A_{ \ B}\hE_{\mu}^{\ B}  \ , \quad
F_{\mu}^{\ A}=\Lambda(l)^A_{ \ B}\hF_{\mu}^{\ B} \ ,
\nonumber \\
\end{eqnarray}
where
\begin{eqnarray}
\hE_{\mu}^{\ A}&=& \left(\begin{array}{cc} N & N^ie_i^{ \ a} \\
0 & e_i^{ \ a} \\ \end{array}\right) \ , \quad  \hF_{\mu}^{\ A}
=\left(\begin{array}{cc} M & M^i f_i^{ \ a} \\
0 & f_i^{ \ a} \\ \end{array}\right) \ ,  \nonumber \\
\hE^\mu_{ \ A}&=& \left(\begin{array}{cc} \frac{1}{N} & 0 \\
-\frac{N^i}{N} & e^i_{ \ a} \\ \end{array}\right) \ , \quad
\hF^\mu_{ \ A}= \left(\begin{array}{cc} \frac{1}{M} & 0 \\
-\frac{M^i}{M} & f^i_{ \ a} \\ \end{array}\right) \ , \nonumber \\
\end{eqnarray}
where $g_{ij}=e_i^{ \ a}e_j^{ \ b}\delta_{ab} \ , f_{ij}=f_i^{ \
a}f_j^{ \ b}\delta_{ab}$.

To proceed further we use the fact that bi-gravity is invariant
under diagonal local Lorentz transformation which implies that we
can partially gauge fix this gauge by imposing $l_a=0$
\cite{Hinterbichler:2012cn}. Note also that since Einstein-Hilbert
actions are invariant under local transformations the action depends
on $p_a$ through the potential term only. Explicitly we find
\begin{eqnarray}
S_1(E^{-1}F)&=&\tr (E^{-1}\hF)=\nonumber \\
&=&
\frac{M}{N}\gamma+ \frac{1}{N} (M^i f_i^{ \ b}p_b -N^i f_i^{ \
b}p_b)+e^i_{ \ a}f_i^{ \ a} +\frac{1}{\gamma+1} (e^j_{ \
a}p^a)(f_j^{ \ b}p_b) \ .
 \nonumber \\
\end{eqnarray}
Using the Hamiltonian analysis performed in Appendix we find
following Hamiltonian
\begin{eqnarray}
H&=&\int d^3\bx (N \mR_0^{(g)}+M \mR_0^{(f)}+ N^i\mR_i^{(g)}+
M^i\mR_i^{(f)}+
\mu^2 Ne\mV+  \nonumber \\
&+&\Lambda^{ab}_{(g)}L^{(g)}_{ab}+\Lambda^{ab}_{(f)}L^{(f)}_{ab}) \
,
\end{eqnarray}
where
\begin{eqnarray}
\mR_0^{(g)}&=&\frac{1}{M_g^2\sqrt{g}}
\pi^{ij}\mG_{ijkl}\pi^{kl}-M_g^2\sqrt{g}R^{(g)} \ , \quad
\mR_0^{(f)}= \frac{1}{M_f^2
\sqrt{f}}\rho^{ij}\tmG_{ijkl}\rho^{kl}-M_f^2\sqrt{f}R^{(f)} \ , \nonumber \\
\mR_i^{(g)}&=&-2g_{ij}\nabla_k\pi^{kj} \ , \quad L_{ab}^{(g)}=
e_{ia}\pi^i_{ \ b}-e_{ib}\pi^i_{ \ a}  \ , \nonumber \\
 \mR_i^{(f)}&=&
-2f_{ij}\tnabla_k\rho^{kj} \ ,  \quad L^{(f)}_{ab}=f_{ia}\rho^i_{ \
b}-f_{ib}\rho^i_{ \ a}  \ ,
 \nonumber \\
\mV&=&\frac{M}{N}\gamma+ \frac{1}{N} (M^i f_i^{ \ b}p_b -N^i f_i^{
\ b}p_b)+e^i_{ \ a}f_i^{ \ a} +\frac{1}{\gamma+1} (e^j_{ \
a}p^a)(f_j^{ \ b}p_b)
 \nonumber \\
\end{eqnarray}
and where $\pi^{ij}$ and $\rho^{ij}$ are momenta conjugate to
$g_{ij}$ and $f_{ij}$ respectively. Further  $\nabla_i$ and
$\tnabla_i$ are covariant derivatives evaluated using the metric
components $g_{ij}$ and $f_{ij}$ respectively. Finally note that
$\mG^{ijkl}$ and $\tmG^{ijkl}$ are de Witt metrics defined as
\begin{equation}
\mG^{ijkl}=\frac{1}{2}(g^{ik}g^{jl}+g^{il}g^{jk})-g^{ij}g^{kl} \ ,
\quad  \tmG^{ijkl}=\frac{1}{2}(f^{ik}f^{jl}+f^{il}f^{jk})-
f^{ij}f^{kl} \
\end{equation}
with inverse
\begin{equation}
\mG_{ijkl}=\frac{1}{2}(g_{ik}g_{jl}+ g_{il}g_{jk})-\frac{1}{2}
g_{ij}g_{kl} \ , \quad \tmG_{ijkl}=\frac{1}{2}(f_{ik}f_{jl}+
f_{il}f_{jk})-\frac{1}{2} f_{ij}f_{kl} \
\end{equation}
that obey the relation
\begin{equation}
\mG_{ijkl}\mG^{klmn}=\frac{1}{2}(\delta_i^m\delta_j^n+
\delta_i^n\delta_j^m)  \ , \quad
\tmG_{ijkl}\tmG^{klmn}=\frac{1}{2}(\delta_i^m\delta_j^n+
\delta_i^n\delta_j^m)  \ .
\end{equation}
   Also note that $e\equiv \det e$. We have also included the primary constraints
$L_{ab}^{(g)}\approx 0 \ ,L_{ab}^{(f)}\approx 0$ into definition of
the Hamiltonian.

An important point is to identify four constraints that are
generators  of the
 diagonal diffeomorphism\footnote{Now due to the specific form of the
 interaction term we have the action that is linear in $N$ and $M$
 and hence the first guess would be that given constraints arise as
 the linear combinations of $\mR_0^{(g)},\mR_0^{(f)}$ and
$\mR_i^{(g)}$ and $\mR_i^{(f)}$. We checked this possibility however
we found that it does not work due to the presence of the constraint
$k^a\approx 0$ defined below. The requirement of the preservation of
the constraints $k^a\approx 0$  led to the secondary constraints
that were functions of $N$ and $M$. Then it was very difficult to
identify four first class constraints that are generators of
diagonal diffeomorphism. It turned out that these generators can be
identified very easily using the ansatz introduced in
\cite{Damour:2002ws} even if it was proposed for the
 case of bimetric gravity formulated
using metric variables.}.  In order to do this we proceed as in
\cite{Damour:2002ws} and introduce following variables
\begin{eqnarray}\label{defnewN}
\bN&=&\sqrt{NM} \ , \quad  n=\sqrt{\frac{N}{M}} \ , \quad
\bN^i=\frac{1}{2}
(N^i+M^i) \ , \quad  n^i=\frac{N^i-M^i}{\sqrt{NM}} \ , \nonumber \\
N&=&\bN n \ , \quad  M=\frac{\bN}{n} \ , \quad
M^i=\bN^i-\frac{1}{2}n^i\bN \ , \quad  N^i=\bN^i+\frac{1}{2}n^i\bN
 \ . \nonumber \\
\end{eqnarray}
Note that their  conjugate momenta are the primary constraints of
the theory
\begin{equation}\label{Pbnprim}
\bP\approx 0 \ , \quad  p\approx 0 \ , \quad  P_i\approx 0 \ , \quad
p_i\approx 0 \
\end{equation}
with  following  non-zero Poisson brackets
\begin{eqnarray}\label{canpb}
\pb{\bN(\bx),\bP(\by)}&=&\delta(\bx-\by) \ , \quad
 \pb{n(\bx),p(\by)}=\delta(\bx-\by) \ , \nonumber \\
\pb{\bN^i(\bx),P_j(\by)}&=&\delta^i_j\delta(\bx-\by) \ , \quad
\pb{n^i(\bx),p_j(\by)}=\delta^i_{j}\delta(\bx-\by) \ .
\nonumber \\
\end{eqnarray}
It is also important to stress that the absence of the time
derivative of $p_a$ in the action implies following primary
constraint
\begin{equation}\label{conka}
k^a\approx 0 \ ,
\end{equation}
where $k^a$ is momentum conjugate to $p_a$ with non-zero Poisson
bracket
\begin{equation}
\pb{p_a(\bx),k^b(\by)}=\delta_a^b\delta(\bx-\by) \ .
\end{equation}
We also have to identify the constraints that are generators of the
diagonal spatial rotations of vielbeins $e_i^{ \ a},f_i^{ \ a}$.
These constraints are  given as  the linear combinations of
$L_{ab}^{(g)},L_{ab}^{(f)}$ and $k^a$. Explicitly, we  introduce
following set of the constraints
\begin{equation}
L_{ab}^{diag}\approx 0 \ , \quad L_{ab}^{br}\approx 0 \ , \quad
k^a\approx 0 \ ,
\end{equation}
where
\begin{eqnarray}
L_{ab}^{diag}&=& e_{ia}\pi^i_{ \ b}-e_{ib}\pi^i_{ \
a}+f_{ia}\rho^i_{\ b}-f_{ib}\rho^i_{ \ a}+p_ak_b-p_bk_a
 \  , \nonumber \\
L_{ab}^{br}&=& e_{ia}\pi^i_{ \ b}-e_{ib}\pi^i_{ \ a}-f_{ia}\rho^i_{\
b} +f_{ib}\rho^i_{ \ a}-p_ak_b+p_bk_a \ ,  \nonumber \\
\end{eqnarray}
where $\pi^i_{ \ a},\rho^i_{ \ a}$ are momenta conjugate to $e_i^{ \
a},f_i^{ \ a}$ respectively. Collecting all these terms together we
find following form of the Hamiltonian
 \begin{eqnarray}
H&=&\int d^3\bx
(\bN\bmR+\bN^i\bmR_i+\Lambda^{ab}_{diag}L_{ab}+\nonumber \\
&+&\Lambda^{ab}_{br}L_{ab}^{br}+v_a k^a+v_n p+v^i
p_i+V_{\bN}\bP+V^iP_i) \ , \nonumber \\
\end{eqnarray}
where
\begin{eqnarray}\label{defbmRtmV}
\bmR&=&n\mR_0^{(g)}+\frac{1}{n}\mR_0^{(f)}+
\frac{1}{2}n^i\mR_i^{(g)}
-\frac{1}{2}n^i\mR_i^{(f)}+\nonumber \\
&+&\mu^2 e\tmV \  , \quad   \bmR_i=
\mR_i^{(g)}+\mR_i^{(f)} \ , \nonumber \\
\end{eqnarray}
where
\begin{eqnarray}\label{deftmV}
\tmV=\frac{\gamma}{n}-n^if_i^{ \ a}p_a + ne^i_{ \ a}f_i^{ \
a}+\frac{n}{\gamma+1}(e^i_{ \ a}p^a) (f_i^{ \ b}p_b) \ .
\nonumber \\
\end{eqnarray}
Now we proceed to the analysis of time development of the primary
constraints (\ref{Pbnprim}) and (\ref{conka})
\begin{eqnarray}
\partial_t \bP&=&\pb{\bP,H}=-\bmR\approx 0 \ ,
\nonumber \\
\partial_t P_{i}&=&\pb{P_i,H}=-\bmR_i\approx 0 \ ,
\nonumber \\
\partial_t p&=&\pb{p,H}=-\mR_0^{(g)}+
\frac{1}{n^2}\mR_0^{(f)}-\mu^2 e\frac{\delta \tmV}{\delta n}
\equiv \mG_n \approx 0 \ , \nonumber \\
\partial_t p_i&=&\pb{p_i,H}=
-\frac{1}{2}(\mR_i^{(g)}-\mR_i^{(f)})-\mu^2e\frac{\delta\tmV}{
\delta n^i}\equiv \mS_i\approx 0 \ ,
\nonumber \\
\partial_t k^a&=&\pb{k^a,H}=-\mu^2e\frac{\delta \tmV}{\delta p_a}
\equiv \mK^a\approx 0 \ .  \nonumber \\
\end{eqnarray}
Finally we have to the check the preservation of the constraints
$L_{ab}^{diag}\approx 0 \ , L_{ab}^{br}\approx 0$. Firstly due to
the fact that $\mR_0^{(g)},\mR_0^{(f)},\mR_i^{(g)},\mR_i^{(f)}$ have
vanishing  Poisson brackets  with $L_{ab}^{(g)},L_{ab}^{(f)}$
according to
 (\ref{pbBTTAPP}) we find that they have also vanishing Poisson
 brackets  with
  both $L_{ab}^{diag}$ and
 $L_{ab}^{br}$. Then the non-zero contribution could follow from the
 Poisson bracket between $L_{ab}^{diag},L_{ab}^{br}$ and $\tmV$.
 Now with the help of the following Poisson brackets
\begin{eqnarray}
\pb{L_{ab}^{diag}(\bx),e_i^{ \ c}(\by)}&=& (e_{ib}\delta^c_a-
e_{ia}\delta^c_b)(\bx)\delta(\bx-\by) \ ,
 \nonumber \\
\pb{L_{ab}^{diag}(\bx),e^i_{ \ c}(\by)}&=&(\delta_{ac}e^i_{ \ b}-
\delta_{bc}e^i_{ \ a})(\bx)\delta(\bx-\by) \ , \nonumber \\
\pb{L_{ab}^{diag}(\bx),f_i^{ \ c}(\by)}&=& (f_{ib}\delta^c_a-
f_{ia}\delta^c_b)(\bx)\delta(\bx-\by) \ ,
 \nonumber \\
\pb{L_{ab}^{diag}(\bx),f^i_{ \ c}(\by)}&=&(\delta_{ac}f^i_{ \ b}-
\delta_{bc}f^i_{ \ a})(\bx)\delta(\bx-\by) \ , \nonumber \\
\pb{L_{ab}^{diag}(\bx),p_c(\by)}&=&-(p_a\delta_{bc}- p_b\delta_{ac})
(\bx)\delta(\bx-\by) \ , \nonumber \\
\pb{L_{ab}^{diag}(\bx),p_cp^c(\by)}&=&0 \nonumber \\
\end{eqnarray}
and also
\begin{eqnarray}
\pb{L_{ab}^{br}(\bx),e_i^{ \ c}(\by)}&=& (e_{ib}\delta^c_a-
e_{ia}\delta^c_b)(\bx)\delta(\bx-\by) \ ,
 \nonumber \\
\pb{L_{ab}^{br}(\bx),e^i_{ \ c}(\by)}&=&(\delta_{ac}e^i_{ \ b}-
\delta_{bc}e^i_{ \ a})(\bx)\delta(\bx-\by) \ , \nonumber \\
\pb{L_{ab}^{br}(\bx),f_i^{ \ c}(\by)}&=& -(f_{ib}\delta^c_a-
f_{ia}\delta^c_b)(\bx)\delta(\bx-\by) \ ,
 \nonumber \\
\pb{L_{ab}^{br}(\bx),f^i_{ \ c}(\by)}&=&-(\delta_{ac}f^i_{ \ b}-
\delta_{bc}f^i_{ \ a})(\bx)\delta(\bx-\by) \ , \nonumber \\
\pb{L_{ab}^{br}(\bx),p_c(\by)}&=&(p_a\delta_{bc}- p_b\delta_{ac})
(\bx)\delta(\bx-\by) \ , \nonumber \\
\pb{L_{ab}^{br}(\bx),p_cp^c(\by)}&=&0\nonumber \\
\end{eqnarray}
we find that the constraint $L_{ab}^{diag}\approx 0$ is preserved
during the time evolution of the system while the  requirement of
the preservation of the constraint $L_{ab}^{br}$ implies
\begin{eqnarray}
\partial_t L_{ab}^{br}(\bx)&=&\pb{L_{ab}^{br}(\bx),H}=\nonumber \\
&=&
2\mu^2\bN e n [(e^j_{ \ b}f_{ja}-e^j_{  \ a}f_{jb})
+\frac{1}{\gamma+1} [(p_a e^j_{ \ b}-p_b e^j_{ \ a}) f_j^{ \ d}p_d]
\equiv 2\mu^2 e  \bN n
\mT_{ab}\approx 0 \ , \nonumber \\
\end{eqnarray}
where we introduced new secondary constraints $\mT_{ab}=-\mT_{ba}$
\begin{equation}\label{defmGab}
\mT_{ab}=e^j_{ \ b}f_{ja}-e^j_{  \ a}f_{jb} +\frac{1}{\gamma+1} (p_a
e^j_{ \ b}-p_b e^j_{ \ a}) f_j^{ \ d}p_d \ .
\end{equation}
As we will see below the  existence of these constraints will be
crucial for the consistency of the theory.
\subsection{Calculation of the Poisson brackets between
$\bmR,\bmR_i$} Before we proceed to the analysis of the stability of
all constraints we would like to show that the Poisson brackets
between the constraints $\bmR$ and $\bmR_i$ vanish on the constraint
surface spanned by all constraints.
 To begin with we introduce the smeared form
 of the constraint $\bmR$
\begin{eqnarray}
\bT(N)= \int d^3\bx N\bmR \ . \nonumber \\
\end{eqnarray}
Then using the Poisson brackets given in (\ref{pbBTTAPP}) and
 following  similar analysis
as in case of metric formulations of bigravity we obtain
\cite{Kluson:2013lza,Kluson:2012ps}
\begin{eqnarray}\label{pbbTN}
\pb{\bT_T(N),\bT_T(M)}&=&\frac{1}{2}\bT_S((N\partial_i M-M\partial_i
N)n^2 g^{ij}) +\frac{1}{2}\bT_S((N\partial_i M-M\partial_i
N)\frac{1}{n^2}f^{ij})+\nonumber \\
&+&\frac{1}{4} \bT_S((N\partial_iM-M\partial_i N)n^in^j) -
\nonumber \\
& -& \bG_S((N\partial_i M-M\partial_i N)n^2 g^{ij}) +\bG_S
((N\partial_i M-M\partial_i N)\frac{1}{n^2}f^{ij}) -\nonumber
\\
&-&\frac{1}{2}\bG_T((N\partial_i M-M\partial_i N)n^i n) + \int
d^3\bx
(N\partial_i M-M\partial_iN)\Sigma^i \ , \nonumber \\
\end{eqnarray}
where
\begin{equation}
\bG_T(N)=\int d^3\bx N\mG_n \ , \quad \bG_S(N^i)= \int d^3\bx
N^i\mS_i \ ,
\end{equation}
and  where
\begin{eqnarray}\label{Sigmanon}
\Sigma^i[\tmV] =\gamma\frac{n^i}{n} -ne^i_{\ a}f_j^{ \
a}n^j-\frac{n}{\gamma+1}e^i_{ \ b}p^b n^jf_j^{ \ a}p_a
+n^2 g^{ij}f_j^{ \ a}p_a-\frac{1}{n^2}f^i_{ \ a}p^a \ .  \nonumber \\
\end{eqnarray}
Note also that we used the extended version of the constraint
$\bmR_i$ given in (\ref{defbmRi}) and we omitted terms proportional
to $L_{ab}^{(g)},L_{ab}^{(f)}$ given in (\ref{pbBTTAPP}).

Our goal is to show that $\Sigma^i[\mV]$ vanishes on the constraint
surface. To do this we use the fact that $n^i$ can be expressed from
the constraint $\mK^a$
\begin{eqnarray}\label{ninon}
n^i=f^i_{ \ a}\left(\frac{1}{\gamma n}p^a-n(e^j_{ \ b}p^b)( f_j^{ \
c}p_c)\frac{1}{(1+\gamma)^2\gamma}p^a+ \frac{1}{1+\gamma} (f_j^{ \
a}e^j_{ \ b}p^b+f_j^{ \ b}p_b e^{ja})+H_a\mK^a\right) \ ,
 \nonumber \\
\end{eqnarray}
where $H_a$ are functions that depend on the phase space variables
whose explicit form is not important for us. To proceed further we
use the fact that from the constraint
 $\mT_{ab}$ we derive
\begin{equation}
f^i_{ \ a}f_j^{  \ b}p_b e^{ja}=\frac{1}{\gamma}\left(
e^{ib}p_b+\frac{1}{1+\gamma}f^{ia}p_a (f_k^{ \ d}p_d)(e^k_{ \
b}p^b)\right)-\frac{f^i_{ \ a}}{2\gamma\sqrt{e}\sqrt{f}} \mG^{ab}p_b
\ .
\end{equation}
Inserting this expression into (\ref{ninon}) we find
\begin{equation}
n^i=\frac{1}{\gamma n}f^i_{ \ a}p^a+ \frac{n}{\gamma} e^{ib}p_b \
\end{equation}
up to terms proportional to $\mT_{ab}$ and $\mK^a$. Finally
inserting this result into (\ref{Sigmanon}) and after some
calculations we find the desired result
\begin{equation}\label{SigmamV}
\Sigma^i[\tmV]=F_a\mK^a+G^{ab}\mT_{ab}\approx 0 \ .
\end{equation}
Then collecting (\ref{pbbTN}) together with (\ref{SigmamV}) we find
that the Poisson bracket between $\bmR$ is proportional to the
constraints $\bmR_i,\mG_n,\mS_i,\mK^a,\mT_{ab}$ which means that it
vanishes on the constraint surface. This is very important result.
Note also an importance of the constraints $\mK^a,\mT_{ab}$ for the
closure of the Poisson brackets between $\bmR$.

As the next step we calculate the Poisson brackets with the
constraints $\bmR_i$. However it turns out that it is more
convenient to consider its following extension
\begin{eqnarray}\label{defbmRi}
\bmR_i=\partial_i np+\partial_in^jp_j+
\partial_j(n^jp_i)+\partial_i p_ak^a+\nonumber \\
+\mR_i^{(e)}+\frac{1}{2}\omega_{ \ i}^{a \ b}(e)L_{ab}^{(e)} +
\mR_i^{(f)}+\frac{1}{2}\omega_{ \ i}^{a \ b}(f)L_{ab}^{(f)} \ .
\nonumber \\
\end{eqnarray}
Let us define  its smeared form
\begin{equation}\label{defdiffgen}
\bT_S(N^i)=\int d^3\bx N^i\bmR_i \ .
\end{equation}
Then we find   following Poisson brackets
\begin{eqnarray}\label{defdifffp}
\pb{\bT_S(N^i),e_i^{ \ c}}&=& -\partial_i N^j e_j^{ \
c}-N^j\partial_j
e_i^{ \ c} \ , \nonumber \\
\pb{\bT_S(N^i),e^i_{ \ c}}&=&\partial_j N^i e^j_{ \ c}-
N^j\partial_j
e^i_{ \ c} \ , \nonumber \\
\pb{\bT_S(N^i),f^i_{ \ c}}&=&\partial_j N^i f^j_{ \ c}-
N^j\partial_j
f^i_{ \ c} \ , \nonumber \\
 \pb{\bT_S(N^i),f_i^{ \ c}}&=& -\partial_i N^j f_j^{ \
c}-N^j\partial_j
f_i^{ \ c} \ , \nonumber \\
\pb{\bT_S(N^i),n^i}&=&-N^j\partial_j n^i+\partial_j N^i n^j \ ,
\nonumber \\
\pb{\bT_S(N^i),p_a}&=&-N^i\partial_i p_a \ , \nonumber \\
\pb{\bT_S(N^i),n}&=&-N^i\partial_i n \  \nonumber \\
\end{eqnarray}
which shows that $\bT_S(N^i)$ is the  generator of the diagonal
spatial diffeomorphism.

Now we are ready to proceed to the calculation of  the Poisson
bracket between $\bT_S(N^i)$ defined in (\ref{defdiffgen}) and
$\bT_T(N)$. In fact, using (\ref{defdifffp}) we easily find
\begin{eqnarray}
\pb{\bT_S(N^i), e \tmV} =- N^k\partial_k[e\tmV]-\partial_kN^ke\tmV \
. \nonumber
\\
\end{eqnarray}
Finally we should calculate the Poisson bracket between $\bT_S(N^i)$
and $\mR_i^{(f)},\mR_i^{(g)}$ and $\mR_0^{(g)},\mR_0^{(f)}$. This is
really easy task using the results given in (\ref{pbBTTAPP})
 so that we find
\begin{equation}\label{bTSbTT}
\pb{\bT_S(N^i),\bT_T(N)}=\bT_T(N^i\partial_iM)
\end{equation}
up to the terms proportional to the primary constraints
$L_{ab}^{(g)}\approx 0 \ , L_{ab}^{(f)}\approx 0$. In the same way
we can find that
\begin{equation}\label{bTSbTS}
\pb{\bT_S(N^i),\bT_S(M^j)}=\bT_S((N^i\partial_iM^j-M^j\partial_iN^j))
\ .
\end{equation}
Using these results we are ready to proceed to the analysis of the
stability of  constraints.
\subsection{Analysis of stability of
constraints } In this section we perform the analysis of the
stability of all constraints.
 Note that for the potential $\tmV$ given
in (\ref{deftmV}) the constraints $\mG_n,\mS_i,\mK^a$ have the form
\begin{eqnarray}\label{mGalphanon}
\mG_n&=&-\mR_0^{(g)}+\frac{1}{n^2}\mR_0^{(f)} +\mu^2 e\left(
\frac{\gamma}{n^2}-e^i_{ \ a}f_i^{ \ a}-\frac{1}{1+\gamma} (e^i_{ \
a}p^a)(f_i^{ \ b}p_b)\right) \ , \nonumber \\
 \mS_i&=&-\frac{1}{2}(\mR_i^{(g)}-\mR_i^{(f)})+\mu^2 e f_i^{ \
a}p_a
, \nonumber \\
\mK^a&=&\mu^2 e\left(\frac{p^a}{\gamma n}-n^i f_i^{ \ a}-n (e^j_{ \
c}p^c)(f_j^{ \ b}p_b) \frac{p^a}{(1+\gamma)^2\gamma}
+\frac{n}{1+\gamma} (f_j^{ \ a}e^j_{ \ b}p^b+f_j^{ \ b}p_b
e^{ja})\right) \ , \nonumber \\
\mT_{ab}&=&(e^j_{ \ b}f_{ja}-e^j_{ \ a}f_{jb}) +\frac{1}{1+\gamma}
(p_a e^j_{ \ b}-p_b e^j_{ \ a})f_j^{ \ d}p_d \ .
\nonumber \\
\end{eqnarray}
It turns out that these constraints could be simplified
considerably. First of all we have  following relation
\begin{equation}
\bmR+n^i\mS_i+n\mG_n=\frac{2}{n}(\mR_0^{(f)}+ \mu^2 e\gamma)
\end{equation}
so that we see that we can consider as an independent constraint
following one
\begin{equation}
\mG'_n=\mR_0^{(f)}+\mu^2e\gamma \ .
\end{equation}
In previous section we also found  the relation
\begin{equation}
n^i=\frac{1}{\gamma n}f^i_{ \ a}p^a+ \frac{n}{\gamma}
e^{ib}p_b+H_a\mK^a+G^{ab}\mT_{ab}
\end{equation}
so that it is possible to define new independent constraint $\tmK^i$
\begin{equation}
\tmK^i= n^i-\frac{1}{\gamma n}f^i_{ \ a}p^a- \frac{n}{\gamma}
e^{ib}p_b \ .
\end{equation}
Then we have following set of  independent secondary constraints
$\mG'_n,\tmK^i,\mS_i,\mT_{ab}$ so that the total Hamiltonian has the
form
\begin{eqnarray}\label{HT1}
H_T&=&\int d^3\bx (\bN\bmR+\bN^i\bmR_i+\Lambda^{ab}_{diag}L_{ab}+
V_{\bN}\bP+V^iP_i +\nonumber \\
&+&\Gamma_n\mG'_n+\Omega_i\tmK^i+\Gamma^i\mS_i+\Gamma^{ab}\mT_{ab}+
v_np+v^i p_i+v_a k^a+v^{ab}L_{ab}^{br}) \ .  \nonumber \\
\end{eqnarray}
Now we are ready to proceed to the analysis of the stability of all
constraints.
 We begin with the
constraints $p_i\approx 0$
\begin{eqnarray}
\partial_t p_i=\pb{p_i,H_T}\approx
-\mu^2 e \Omega_i =0 \  \nonumber \\
 \end{eqnarray}
 that has the solution $\Omega_i=0$, where $\Omega_i$ is Lagrange
 multiplier corresponding the constraint $\tmK^i$. In case
 of $p$ we find
\begin{equation}
\partial_t p=\pb{p,H_T}\approx 0 \ .
\end{equation}
 Further, the requirement of
the preservation of $L_{ab}^{br}$ takes the form
\begin{equation}\label{preserLabbr1}
\partial_t L_{ab}^{br}=\pb{L_{ab}^{br},H_T}=
\Gamma^{cd}\triangle_{L_{ab}^{br},\mG_{cd}}=0 \ ,
\end{equation}
where
$\pb{L_{ab}^{br}(\bx),\mG_{cd}(\by)}=\triangle_{L_{ab}^{br},\mG_{cd}}
(\bx)\delta(\bx-\by)$ and where we used the fact that $\Omega_i=0$
together with
\begin{equation}
\pb{L_{ab}^{br}(\bx), \mG'_n(\by)}=\pb{L_{ab}^{br}(\bx),\mS_i(\by)}=
0 \ .
\end{equation}
Then it can be explicitly checked that
 the matrix $\triangle_{L_{ab}^{br},\mG_{cd}}$
is non-singular and hence the solution of (\ref{preserLabbr1}) is $
\Gamma^{cd}=0$.

Using these results it is easy to analyze the requirement of the
preservation of the constraints  $k^a\approx 0$
\begin{equation}
\partial_t k^a=\pb{k^a,H_T}\approx
-\mu^2 e \frac{p^a}{\gamma} -\Gamma^i\mu^2 e f_i^{ \ a} =0 \
\end{equation}
so that we obtain
\begin{equation}
\Gamma^i=-\frac{p^a}{\gamma}f^i_{ \ a} \Gamma_n \ .
\end{equation}
This result however suggests to consider as an  independent
constraint following one
\begin{equation}
\tmG_n= \mR_0^{(f)}+\mu^2e\gamma -\frac{1}{\gamma}\mS_i f^i_{ \
a}p^a \
\end{equation}
and following  total Hamiltonian
\begin{eqnarray}
H_T&=&\int d^3\bx (\bN\bmR+\bN^i\bmR_i + \Lambda^{ab}_{diag}L_{ab}+
V_{\bN}\bP+V^iP_i +\nonumber \\
&+&\Gamma^i\mS_i+\Omega_i\tmK^i+\Gamma_n\tmG_n+ v^n p+ v^i p_i+ v_a
k^a+v^{ab}L_{ab}^{br}+\Gamma^{ab}\mT_{ab}) \ .  \nonumber \\
\end{eqnarray}
Repeating the analysis as above we find that $p$ is trivially
preserved and also $\Omega_i=\Gamma^{ab}=0$.
Further,  the time evolution of the constraint $k^a\approx 0$ is
given by equation
\begin{eqnarray}
\partial_t k^a=\pb{k^a,H_T}\approx
-\Gamma^i\mu^2 e f_i^{ \ a}=0 \nonumber \\
\end{eqnarray}
that due to the fact that $f_i^{ \ a}$ is non-singular implies that
$\Gamma^i=0$. Finally it is also clear that $\bP,P_i$ are trivially
preserved.

Now we proceed to the analysis of the time evolution of the
constraint $\tmG_n,\mS_i,\tmK^i$ together with $\bmR$ and $\bmR_i$.
First of all it is easy to see that  that the secondary constraints
$\tmG_n,\mS_i,\tmK^i,\tmG_{ab}$  are invariant under diagonal
spatial diffeomorphism. Then with the help of (\ref{bTSbTT}) and
(\ref{bTSbTS}) we find that $\bmR_i$ are preserved during the time
evolution of the system.

More interesting situation occurs in case of the time evolution of
the constraints $\tmG_n$ and $\bmR$ which is mainly determined by
following Poisson bracket
\begin{equation}\label{btTtmG}
\pb{\bT_T(N),\int d^3\bx M\tmG_n}\approx \int d^3\bx N(\bx)M(\bx)
\tmG_n^{II}(\bx) \ ,
\end{equation}
where
\begin{equation}
\tmG_n^{II}=\tnabla_i
(f^{ij}\mR_j^{(f)})+\frac{1}{2}n^i\mu^2\partial_i \gamma e +\dots \
,
\end{equation}
and  where $\dots$ means other terms that depend on phase space
variables. Note that the explicit form of  $\tmG_n^{II}$, which is
very complicated,  is not important for us. However it is crucial
and non-trivial fact that the Poisson bracket (\ref{btTtmG}) does
not contain terms proportional to $M\partial_iN$ or $M\partial_iN$.
Then the local form (\ref{btTtmG}) has the form
\begin{equation}\label{locbtTtmG}
\pb{\bmR(\bx),\tmG_n(\by)}\approx \tmG^{II}_n(\bx)\delta(\bx-\by) \
\end{equation}
so that there are no derivative of the delta function on the right
side of the previous equation. This fact is very important for the
consistency of given theory.

 Now we are
ready to proceed to the analysis of the consistency of the secondary
constraints. In case of $\tmG_n$ we obtain
\begin{eqnarray}\label{parttmGn}
\partial_t \tmG_n(\bx)=\pb{\tmG_n(\bx),H_T}\approx
-\bN(\bx)\tmG^{II}_n(\bx)
\end{eqnarray}
using the fact that the Poisson bracket
$\pb{\tmG_n(\bx),\tmG_n(\by)} $ is weakly zero as follows from
\begin{eqnarray}\label{pbmGnn}
& &\pb{\int d^3\bx N\tmG_n(\bx),\int d^3\by M\tmG_n(\by)}=
\nonumber \\
&=&\frac{1}{2}\bT_S((N\partial_i M-M\partial_i N)f^{ij}) +
\frac{1}{4} \bT_S\left(\frac{1}{\gamma^2}f^i_{ \ a}p^a f^j_{ \ b}p^b
(\partial_i
NM-M\partial_iN)\right)+ \nonumber \\
 &+& \int d^3\bx
(N\partial_i M-M\partial_iN)f^{ij}\mS_j-\nonumber \\
&-&\frac{1}{2}\int d^3\bx (N\partial_i M-M\partial_iN)\frac{f^i_{ \
a}p^2}{\gamma}\mG_n'\approx 0 \ . \nonumber \\
\end{eqnarray}
 On the other hand the time evolution of the constraint
$\bmR$ is equal to
\begin{eqnarray}\label{partbmRT}
\partial_t\bmR(\bx)=\pb{\bmR(\bx),H_T}\approx
\Gamma_n \tmG_n^{II}(\bx)=0 \nonumber \\
\end{eqnarray}
using (\ref{pbbTN}),(\ref{SigmamV}) and (\ref{bTSbTT}) together with
the fact that $\Gamma^i=\Gamma^{ab}=\Omega_i=0$.  Now it is crucial
to find non-trivial solution of (\ref{partbmRT}). In case when
$\tmG_n^{II}$ were constant on the whole phase space we would find
that the only possible solution is $\Gamma_n=0$. Then from
(\ref{parttmGn}) we would also find $N=0$ and hence we should
interpret $\bmR$ together with $\tmG_n$ as the second class
constraints. However this is very unsatisfactory result since it
would imply the lack of the Hamiltonian constraint while the theory
is manifestly invariant under diagonal diffeomorphism. Fortunately
$\tmG_n^{II}$ depends on the phase space variables so that it is
more natural to obey  (\ref{partbmRT}) when we say that
$\tmG_n^{II}$ is an additional constraint imposed on the system.

Now with this interpretation we find that (\ref{parttmGn}) vanishes
on the constraint surface when $\Gamma_n=0$.
As the next step we will analyze the requirement of the preservation
of the constraints $\mS_i,\mT_{ab},\tmK^i$ which however simplifies
considerably due to the fact that
$\Gamma_n=\Gamma^i=\Gamma^{ab}=\Omega_i=0$. We start with the
constraint $\mS_i$
\begin{eqnarray}\label{detva}
\partial_t\mS_i=\pb{\mS_i,H_T}\approx
\int d^3\bx\bN(\bx)\pb{\mS_i,\bmR(\bx)}+f_i^{ \ a}v_a=0 \ , \nonumber \\
\end{eqnarray}
using also the fact that $\pb{\mS_i(\bx),L_{ab}^{br}(\by)}=0$. Now
due to the fact that the matrix $f_i^{ \ a}$ is non-singular we find
that this equation  can be solved for $v_a$.

In case of the constraints $\mT_{ab}$ we find
\begin{eqnarray}\label{partmgabnon}
\partial_t\mT_{ab}=\pb{\mT_{ab},H_T}
\approx \int d^3\bx \bN(\bx)\pb{\mT_{ab},\bmR(\bx)}+
 \triangle_{\mT_{ab},k^c}v_c+ \triangle_{\mT_{ab},L^{br}_{cd}}
v^{cd}=0 \ ,
 \nonumber \\
\end{eqnarray}
where the matrix $\triangle_{\mT_{ab},k^c}$ is  defined as
\begin{eqnarray}
\pb{\mT_{ab}(\bx),k^c(\by)}\equiv
\triangle_{\mT_{ab},k^c}(\bx)\delta(\bx-\by) \ . \nonumber \\
\end{eqnarray}
Now using the fact that the matrix
$\triangle_{\mT_{ab},L^{br}_{cd}}$ is non-singular and that $v_a$
were determined by (\ref{detva}) we find that the equation
(\ref{partmgabnon}) can be explicitly solved for $v^{ab}$.

Finally we proceed to the analysis of the equation of motion of the
constraint $\tmK^i$
\begin{eqnarray}\label{parttmGanon}
\partial_t \tmK^i&=&\pb{\tmK^i,H_T}\approx
\int d^3\bx \bN(\bx)\pb{\tmK^i,\bmR(\bx)}+ v^n\triangle_{\tmK^i,p}+
\nonumber \\
&+&v^{ab}\triangle_{\tmK^i,L_{ab}^{br}}+ v^i + v_c
\triangle_{\tmK^i,k^c}=0 \ ,  \nonumber \\
\end{eqnarray}
where we defined
\begin{eqnarray}
\pb{\tmK^i(\bx),p(\by)}&=&\left[-\frac{1}{\gamma n^2}f^i_{ \ a}p^a
+\frac{1}{\gamma}e^{ib}p_b\right]\delta(\bx-\by)\equiv
\triangle_{\tmK^i,p}(\bx)\delta(\bx-\by) \ , \nonumber \\
\pb{\tmK^i(\bx),L_{ab}^{br}(\by)}&=&\frac{2n}{\gamma}(e^i_{ \ a}p_b-
e^i_{ \ b}p_a)(\bx)\delta(\bx-\by)\equiv
\triangle_{\tmK^i,L_{ab}^{br}}\delta(\bx-\by) \ , \nonumber \\
\pb{\tmK^i(\bx),k^c(\by)}&=&
\frac{1}{\gamma^2}p^c(\tmK^i-n^i)\delta(\bx-\by)- \frac{1}{\gamma}
(\frac{1}{n}f^i_{ \ c}+n e^i_{ \ c})\delta(\bx-\by)\approx \nonumber
\\
&\approx & -\left(\frac{1}{\gamma^2}p^c n^i+ \frac{1}{\gamma}
\frac{1}{n}f^i_{ \ c}+\frac{n}{\gamma} e^i_{ \
c}\right)\delta(\bx-\by) \equiv
\triangle_{\tmK^i,k^c}(\bx)\delta(\bx-\by) \ . \nonumber \\
\end{eqnarray}
We see that  (\ref{parttmGanon}) can be solved for $v^i$ knowing the
Lagrange multipliers $v_{ab},v_c,v_n$. Note that $v_n$ is still
undetermined which is the reflection of the fact that $p\approx 0$
is the first class constraint.

Finally we should check the stability of all constraints with the
constraint $\tmG_n^{II}\approx 0$ included. However it turns out
that there is non-zero Poisson bracket between $\tmG_n^{II}\approx
0$ and $\tmG_n\approx 0$ and these are the second class constraints.
Then the analysis of the stability of all constraints is the same as
above and we will not repeat here \footnote{It is necessary to
stress one important point. From the form of the constraint
$\tmG_n^{II}$ we find that $\pb{\mG_n^{II},\bmR}\neq 0$ and hence we
could say that $\bmR$ is second class constraint. Clearly this is
rather unsatisfactory result since we would have three second class
constraints while we should expect two second class constraints and
one first class constraint. In order to see how to resolve this
puzzle let us consider the case where the constraints
$\bmR,\tmG_n,\tmG_n^{II}$ do not depend on spatial coordinates
keeping in mind that extension of this analysis to the more general
case is straightforward. Note that we can use this approximation
since we know that $\pb{\tmG_n(\bx),\tmG_n(\by)}\approx 0$ as
follows from (\ref{pbmGnn}).
  Then the requirement of the stability of
the constraints $\bmR,\tmG_n,\tmG_n^{II}$ implies following
equations $\partial_t\bmR\approx
\pb{\bmR,\tmG^{II}_n}\lambda^{II}_n=0$ that has solution
$\lambda^{II}_n=0$. Then we also find that the constraint $\tmG_n$
is preserved. Finally the requirement of the stability of the
constraint $\tmG_n^{II}$ implies following equation
$\partial_t\tmG^{II}\approx \pb{\tmG^{II},\bmR}N+ \lambda^I_n
\pb{\tmG^{II}_n,\tmG^I_n}=0$ that can be schematically solved as
$\lambda^I=-\frac{\pb{\tmG^{II}_n,\bmR}}{\pb{\mG^{II}_n,\mG^I_n}}N$.
Then it is natural to define new Hamiltonian constraint $\bmR'=
\bmR-\frac{\pb{\tmG^{II}_n,\bmR}}{\pb{\tmG^{II}_n,\tmG^I_n}}\tmG^I_n$.
Now it ie easy to see that
$\pb{\tmG_n,\bmR'}=\pb{\tmG^{II}_n,\bmR'}=0$ and hence we find one
first class constraint $\bmR'$ and two second class constraints
$\tmG_n,\tmG_n^{II}$ as expected.}. Let us outline our results and
determine the physical degrees of freedom of given theory. We have
$N_{f.c.c.}=12$ first class constraints
$\bmR,\bmR_i,\bP,P_i,L_{ab}^{diag},p$. Then we have $N_{s.c.c.}=20$
second class constraints
$p_i,k^a,L_{ab}^{br},\tmG_n,\tmG_n^{II},\mS_i,\tmK^i,\mT_{ab}$. We
also have $N_{ph.s.d.f.}=58$ phase space degrees of freedom
$\bN,\bP,\bN^i,P_i, n,p,n^i,p_i,p_a,k^a,e_i^{ \ a},\pi^i_{ \
a},f_i^{ \ a},\rho^i_{ \ a}$. Then the number of physical degrees of
freedom $N_{p.d.f.}$ is \cite{Henneaux:1992ig}
\begin{equation}
N_{p.d.f.}=N_{ph.s.d.f.}-2N_{f.c.c.}-N_{s.c.c.}=14
\end{equation}
that could be interpreted as $4$ physical degrees of freedom of the
massless graviton, $10$ physical degrees of freedom corresponding to
the massive graviton.  In other words we have shown that the
bi-gravity  in the vierbein formulation is ghost free.

\section{Conclusion}\label{third}
This paper was devoted to the Hamiltonian analysis of the bimetric
theory of gravity in the form introduced in
\cite{Hinterbichler:2012cn}. We found corresponding Hamiltonian and
determined the primary constraints of the theory. Then we analyzed
the requirement of the preservation of these constraints and we
determined corresponding secondary constraints. Finally we
determined conditions when these constraints are preserved and we
found that there is an additional constraint. As a result the
constraint structure of given theory suggests that this theory is
free of ghosts.

However   it is still  important to stress that even if the
non-linear massive gravity is  ghost free this does not mean that
given theory is consistent. In fact, it was shown that non-linear
massive gravity suffers from the superluminality in its decoupling
limit \cite{Gruzinov:2011sq,Burrage:2011cr,deFromont:2013iwa}. It
was also shown that generally contain the tachyonic modes
\cite{Deser:2001dt,Deser:2013uy}. Further, the analysis of
cosmological properties of non-linear massive gravity showed that it
exhibits the ghost instabilities about its homogeneous solutions
\cite{DeFelice:2012mx,DeFelice:2013awa,DeFelice:2013bxa}, see also
\cite{Berezhiani:2013dw,Berezhiani:2013dca} \footnote{However quiet
recently the improved version of non-linear massive gravity was
proposed in \cite{DeFelice:2013tsa} that is claimed to be unitary
with all degrees of freedom propagating on a homogeneous, isotropic
and self-accelerating de Sitter background.}. On the other hand it
was shown very recently in \cite{DeFelice:2014nja} that non-linear
bimetric theory of gravity could lead to viable cosmology under some
conditions. In fact, the bimetric theory of gravity has one
important advantage with respect to non-linear massive gravity when
the second metric is not fixed by hand but it is dynamical as well.
Clearly bimetric theory of gravity is very promising generalization
of gravity that deserves to be studied further.


 {\bf Acknowledgement:}
\\
I would like to thank S. Alexandrov for very useful discussions and
for his finding of the crucial error in the first version of this
paper.
 This work   was
supported by the Grant agency of the Czech republic under the grant
P201/12/G028. \vskip 5mm

\begin{appendix}
\section{Appendix: Hamiltonian Formalism of General Relativity in Vierbein
Formulaton}\label{appendix} In this Appendix we  perform the
Hamiltonian formalism of the general relativity in vierbein
formulation. We mostly follow
\cite{Peldan:1993hi,Nicolai:1992xx,Henneaux:1983vi,Charap:1988vz}.

 Let us consider the general relativity Lagrangian density written in    the form
\begin{equation}
\mL=M_p^2 \det E(\Omega_A^{ \ AB}\Omega_{CB}^{ \ C}
-\frac{1}{2}\Omega^{CAB}\Omega_{CBA}
-\frac{1}{4}\Omega^{BAC}\Omega_{BAC}) \ ,
\end{equation}
where
\begin{equation}
\Omega^{CAB}=E^{\mu C}E^{\nu A}\partial_{[\mu }E^B_{ \ \nu]} \ ,
E^{\mu A}=E^\mu_{ \ B}\eta^{BA} \ ,
\end{equation}
and where $\mu,\nu,\dots=0,1,2,3$ and where $A,B,\dots=0,1,2,3$.
Note that by definition we have two covariant derivatives $\hmD_\mu$
and $\hnabla_\mu$. $\hmD_\mu$ is covariant with respect to both
general coordinate transformations in spacetime as well as local
Lorentz transformations on the flat index while $\hnabla_\mu$ is
covariant under general coordinate transformations. We have
\begin{eqnarray}
\hmD_\mu \lambda^{\nu A}&=&\partial_\mu \lambda^{\nu A}+
\hat{\Gamma}^\nu_{\mu\gamma}\lambda^{\gamma A}+ \hat{\omega}^{ \
A}_{\mu \ B}
\lambda^{\nu B} \ , \nonumber \\
\hnabla_\mu\lambda^{\nu A}&=&
\partial_\mu \lambda^{\nu A}+\hat{\Gamma}^\nu_{\alpha \gamma}
\lambda^{\gamma A} \ .  \nonumber \\
\end{eqnarray}
We require that these covariant derivatives are compatible with the
vierbein  and the metric
\begin{eqnarray}\label{defcom}
\hmD_\mu E_\nu^{ \ A}&=&0 \ , \nonumber \\
\hnabla_\mu \hg_{\nu\sigma}&=&\hmD_\mu \hg_{\nu\sigma}=0 \ .  \nonumber \\
\end{eqnarray}
where $\hg_{\mu\nu}=E_\mu^{ \ A}E_{\nu}^{ \ B}\eta_{AB}$. Note that
from (\ref{defcom}) we obtain
\begin{equation}
\hnabla_\mu E^\nu_{ \ A}= -\hat{\omega}_{\mu A}^{\quad B}E^\nu_{ \
B} \ .
\end{equation}
From (\ref{defcom}) and requiring $\hat{\Gamma}_{[\mu\nu]}^\rho=0$
it is possible to uniquely determine $\hat{\Gamma}_{\mu\nu}^\rho$
and $\hat{\omega}^{\ A}_{\mu \ B}$ as functions of vierbein $E_\mu^{
\ A}$. Explicitly, the first equation in (\ref{defcom}) can be
solved as
\begin{equation}
\hat{\omega}_\mu^{ \ AB}= \frac{1}{2}e_{\mu C}(\Omega^{CAB}+
\Omega^{BCA} -\Omega^{ABC} ) \ .
\end{equation}
Let us now consider following $3+1$ decomposition of tetrad
\begin{equation}\label{E31}
E_0^{ \ A}=NN^A+N^aV_a^{ \ A} \ , E_i^{ \ A}=V_i^{ \ A} \ ,
N^B\eta_{AB}V_i^{ \ B}=0 \ , N^A\eta_{AB}N^B=-1 \ ,
\end{equation}
where  $i,j,k,\dots=1,2,3$ and $a,b,c,\dots=1,2,3$. The inverse
vielbein obeys
\begin{equation}
E^\mu_{ \ B}E_\mu^{ \ C}=\delta_B^{ \ C} \ , E^\mu_{ \ A}E_\nu^{ \
A}=\delta^\mu_{\ \nu} \ .
\end{equation}
Using this decomposition it is rather straightforward perform the
Legendre transform using this decomposition. However it is more
convenient to partly break the manifest Lorentz invariance in such a
way that the vierbein takes the upper triangular form (\ref{utf}).
In this case we identify $V_i^{ \ a}$ with $e_i^{ \ a}$ where $e_i^{
\ a}$ defines the three dimensional metric $g_{ij}=e_i^{ \ a}e_j^{ \
b}\delta_{ab}$.

 Now using (\ref{E31}) and also the partial gauge fixing we obtain
 following decomposition of $\Omega^{ABC}$
\begin{eqnarray}
\Omega^{0 \ b}_{\ a}&=& \frac{1}{N} e^i_{ \ a}(\partial_0e_i^{ \ b}-
N^j \partial_j e_i^{ \ b}-e^i_{ \ a}e_j^{ \ b}\partial_i N^j) \ ,
\nonumber \\
\Omega^{0 \ 0}_{\ a}&=&\frac{1}{N}e^i_{ \ a}\partial_i N \ ,
\nonumber
\\
\Omega^{a \ 0}_{ \ b}&=&0 \ , \nonumber \\
\Omega_{ab}^{\ \ c}&=& e^i_{ \ a}e^j_{ \ b} (\partial_i
e_j^{ \ c}-\partial_j e_i^{ \ c}) \ .  \nonumber \\
\end{eqnarray}
The general relativity Lagrangian  now takes the form
\begin{eqnarray}\label{mLvier}
\mL&=&M_P^2Ne\left(-\frac{1}{2}\Omega^{0(ab)}\Omega_{0ab}+\Omega^{0\
a}_{ \ a}\Omega_{0b}^{ \ \  b}+2\Omega^{a \  0}_{ \ 0}
\Omega_{ab}^{ \ \  b}\right.-\nonumber \\
& & \left.-\frac{1}{4}\Omega^{abc}\Omega_{abc}-\frac{1}{2}
\Omega^{abc}\Omega_{acb}+ \Omega_c^{ \  ac} \Omega_{ba}^{ \ \
 b}\right) \nonumber \\
\end{eqnarray}
where $e=\det e_i^{ \ a}$ and where we have following convention
\begin{eqnarray}\label{conX}
X^{(ab)}=X^{ab}+X^{ba} \ ,
X^{(ab)}X_{(ab)}=
 2X^{(ab)}X_{ba} \ . \nonumber \\
\end{eqnarray}
Note that we can write
\begin{eqnarray}
N e{}^{(3)}R=N e(\Omega_b^{ \ ab}\Omega_{ca}^{ \ \  c} -\frac{1}{4}
\Omega^{abc}\Omega_{abc}-\frac{1}{2}\Omega^{abc}\Omega_{acb})+
\nonumber \\
+\nabla_{[i}(N e e^{ja}\nabla_{j]}e^i_{ \ a}) -2e e^{i}_{ \
a}\partial_{[i}e_{j]}^{ \ a} g^{jk}\partial_k N \ ,
 \nonumber \\
\end{eqnarray}
where $\nabla_i$ is covariant derivative compatible with $g_{ij}$ so
that $\nabla_i g_{jk}=0$. Then  neglecting the surface term we find
that (\ref{mLvier}) has the form that is suitable for the
Hamiltonian formulation
\begin{equation}\label{mLvierham}
\mL=M_g^2Ne\left(-\frac{1}{2}\Omega^{0ab}\Omega_{0(ab)} +\Omega^{0 \
a}_{ \  a} \Omega_{0b}^{ \  \ b}+{}^{(3)}R\right) \ ,
\end{equation}
where
\begin{eqnarray}
\Omega^{0ab}=\delta^{ac}\Omega^{0 \ b}_{\ c} \ , \quad \Omega_{0ab}=
 -\Omega^{0 \ d}_{ \
b}\delta_{dc} \ , \quad  \Omega_{0b}^{ \ \  b}=
-\Omega^{0 \ b}_{ \ b} \ . \nonumber \\
\end{eqnarray}
 Note that (\ref{conX}) implies
\begin{equation}
\Omega^{0(ab)}\Omega_{0ab}=\frac{1}{2} \Omega^{0(ab)}\Omega_{0(ab)}
\ .
\end{equation}
Then from (\ref{mLvierham}) we find the momenta $\pi^i_{ \ a}$
conjugate to $e_i^{ \ a}$
\begin{equation}\label{pidef}
\pi^i_{ \ a}=\frac{\delta \mL}{\delta \partial_0e_i^{ \ a}}=M_P^2 e(
e^i_{ \ c}\Omega^{0(cd)}\delta_{da}-2\Omega^{0 \ b}_{ \ b} e^i_{ \
a})
\end{equation}
so that
\begin{equation}
\Omega^{0 \ a}_{\ a}=-\frac{1}{4e}e_i^{ \ a}\pi^i_{ \ a} \ .
\end{equation}
Then it is easy to express $\Omega^{0(ab)}$ as function of $\pi^i_{
\ a}$ and $e_i^{ \ a}$
\begin{equation}
\Omega^{0(ab)}=\frac{1}{e}(e_i^{ \ a}\pi^i_{ \ c}\delta^{cb}-
\frac{1}{2}\delta^{ab}e_i^{ \ c}\pi^i_{ \ c}) \ .
\end{equation}
Using this result we easily find the Hamiltonian from
(\ref{mLvierham})
and hence corresponding Hamiltonian
\begin{eqnarray}\label{defH}
\mH=
N\mR+N^i \mR_i \ ,
\nonumber \\
\end{eqnarray}
where
\begin{eqnarray}
\mR
&=&\frac{1}{4M_g^2e}(e_i^{ \ a}\pi^i_{ \ c}\delta^{cg} \delta_{ae}
e_j^{ \ e}\pi^j_{ \ g}-\frac{1}{2}(e_i^{ \ a}\pi^i_{ \
a})^2)-M_g^2e{}^{(3)}R \ , \nonumber \\
\mR_i&=&-e_i^{ \ b}\mD_j\pi^j_{ \ b} \ , \nonumber \\
\end{eqnarray}
where $\mD_i$ is covariant derivative compatible with $e_i^{ \ a}$
\begin{equation}
\mD_i e_j^{ \ a}=0 \ .
\end{equation}
Note also that we  neglected the total derivative terms in the
Hamiltonian (\ref{defH}). It is also important to stress that
(\ref{pidef}) implies following primary constraints
\begin{equation}\label{defLab}
L_{ab}=e_{ia}\pi^i_{ \ b}-e_{ib}\pi^i_{ \ a} \approx 0
\end{equation}
By definition $L_{ab}$ are antisymmetric so that there are three
constraints $L_{ab}$ in three dimensions.

 To proceed further we
need an explicit form of $\mD_i \pi^j_{ \ a}$. Since $\pi^i_{ \ a}$
is the density of weight one we have \footnote{ Note that the spin
connection has following prescription when it acts on object with
upper and lower Lorentz indices
\begin{equation}
\mD_i X^a_{ \ b}=\partial_i X^a_{ \ b}+\omega^{\ a}_{i \ c}X^c_{ \
b}-\omega^{ \ c}_{i \ b}X^a_{ \ c}
\end{equation} \ . }
\begin{eqnarray}
\mD_i\pi^j_{ \ a}=\partial_i\pi^j_{ \ a}+\Gamma^j_{ik}\pi^k_{ \ a}
-\Gamma^m_{mi}\pi^j_{ \ a}-\omega_{i\ a}^{ \   b}\pi^j_{ \ b}
\nonumber \\
\end{eqnarray}
so that \begin{equation}
\mD_i\pi^i_{ \ a}=
\partial_i\pi^i_{ \ a}+\omega_{i \ a}^{ \  b}\pi^i_{ \ b}
\end{equation}
It is also convenient to introduce the notation
\begin{equation}\label{defpiij}
\pi^{ij}=\frac{1}{4}(\pi^i_{ \ a}e^{ja}+ e^{ia}\pi^j_{ \ a})
\end{equation}
that it is
 similar as the notation  used in \cite{Henneaux:1983vi}. Then it
 is easy to see that the Hamiltonian constraint $\mR$ takes the
 familiar form
\begin{equation}
\mR=\frac{1}{\sqrt{g}M_g^2}\mG_{ijkl}\pi^{ij}\pi^{kl}-M_g^2\sqrt{g}{}^{(3)}R
\ ,
\end{equation}
where
\begin{equation}
\mG_{ijkl}=\frac{1}{2}(g_{ik}g_{jl}+g_{il}g_{jk}-g_{ij}g_{kl}) \ ,
\quad
\mG^{ijkl}=\frac{1}{2}(g^{ik}g^{jl}+g^{il}g^{jk})-g^{ij}g^{kl} \ ,
\end{equation}
that obey the relation
\begin{equation}
\mG_{ijkl}\mG^{klmn}=\frac{1}{2}(\delta_i^m\delta_j^n+
\delta_i^n\delta_j^m) \ .
\end{equation}
Note also that in the same way we can write
\begin{equation}
\mR_i=-e_i^{ \ a}\mD_j\pi^j_{ \ b}=
 -2\nabla_i \pi^i_{ \
j}
\end{equation}
using the fact that
\begin{eqnarray}
 -2\nabla_i \pi^i_{ \
j}\approx -\nabla_i\pi^i_{ \ a}e_j^{ \ a}
-\pi^i_{ \ a}\nabla_i e_j^{ \ a}= \nonumber \\
=-\nabla_i\pi^i_{ \ a}a_j^{ \ a}+\pi^i_{ \ a}\omega^{ \ a}_{i \
b}e_j^{ \ b}=-e_i^{ \ a}\mD_j\pi^j_{ \ b}\nonumber \\
\end{eqnarray}
using also the fact that
\begin{eqnarray}
\pi^i_{ \ j}=\pi^{ik}g_{kj}= \frac{1}{2} \pi^i_{ \ a}e_j^{ \ a}
+L_{ad}e^{ia}e_j^{ \ d} \ .  \nonumber \\
\end{eqnarray}
By definition the canonical variables are $e_i^{ \ a}$ and $\pi^j_{
\ b}$ with following canonical Poisson brackets
\begin{equation}\label{canPB}
\pb{e_i^{ \ a}(\bx),\pi^j_{ \ b}(\by)}= \delta_i^j \delta^a_b
\delta(\bx-\by) \
\end{equation}
so that we obtain
\begin{eqnarray}
\pb{g_{ij}(\bx),\pi^{kl}(\by)}=\frac{1}{2}
(\delta_i^k\delta_j^l+\delta_i^l\delta_j^k)
\delta(\bx-\by) \nonumber \\
\end{eqnarray}
On the other hand from (\ref{defpiij}) and from (\ref{canPB}) we
find
\begin{eqnarray}
\pb{\pi^{ij}(\bx),\pi^{kl}(\by)}&=&\frac{1}{16}(g^{il}L^{kj}+g^{jl}L^{ki}+g^{jk}L^{li}+
g^{ik}L^{lj})\delta(\bx-\by)= \nonumber \\
&=&\mu^{ijkl}\delta(\bx-\by) \ . \nonumber \\
\end{eqnarray}
This result implies that there are additional terms when we
calculate the Poisson brackets between the constraints as was nice
shown in \cite{Henneaux:1983vi}. More precisely, let us introduce
the smeared form of the constraints $\mR,\mR_i$ and $L_{ab}$
\begin{equation}\label{defsmearcon}
\bT_T(N)=\int d^3\bx N\mR \ , \quad  \bT_S(N^i)=\int d^3\bx N^i
\mR_i \ , \quad  \bL(N^{ab})=\int d^3\bx N^{ab}L_{ab} \ .
\end{equation}
Then, following  \cite{Henneaux:1983vi} we  find
\begin{eqnarray}\label{pbBTTAPP}
\pb{\bT_T(N),\bT_T(M)}&=& \bT_S((N\partial_iM-M\partial_iN)g^{ij}) \
,
\nonumber \\
\pb{\bT_S(N^i),\bT_T(M)}&=&\bT_T(N^i\partial_iM)+\int d^3\bx
\nabla_j
N^i \lambda_i^{ \ j}M \ , \nonumber \\
\pb{\bT_S(N^i),\bT_S(M^j)}&=&\bT_S((N^i\partial_iM^j-M^j\partial_iN^j))+
\int d^3\bx \nabla_k N^i\mu_{i \ j}^{\ k  \ l}\nabla_lM^j \ ,
\nonumber
\\
\pb{\bT_T(N),\bL(N^{ab})}&=&0 \ , \pb{\bT_S(N^i),\bL(N^{ab})}=0 \ ,
\nonumber \\
\pb{L_{ab}(\bx),L_{cd}(\by)}&=&
(\eta_{ad}L_{bc}+\eta_{bc}L_{ad}-\eta_{bd}L_{ac}
-\eta_{ac}L_{bd})\delta(\bx-\by) \ ,  \nonumber \\
\end{eqnarray}
where
\begin{eqnarray}
\lambda^{ij}&=&-4\mu^{ijkl}K_{kl}= -\frac{1}{2} (K^j_{ \ k}L^{ik}-
K^i_{ \ k}L^{jk}) \ , \nonumber \\
K_{ij}&=&\frac{1}{\sqrt{g}}(\frac{1}{2}\pi^{mn}g_{nm}g_{ij}-\pi_{ij})
\ .
\nonumber \\
\end{eqnarray}
We see that there are additional terms on the right side of the
Poisson brackets between constraints that are proportional to the
primary constraints $L_{ab}$. These terms also vanish on the
constraints surface. For that reason we will not write the explicit
form of these terms in the calculations performed in the main body
of the paper.
\end{appendix}



\begin{thebibliography}{20}

\bibitem{Salam:1969rq}
  A.~Salam and J.~A.~Strathdee,
\emph{``Nonlinear realizations. 1: The Role of Goldstone bosons,''}
  Phys.\ Rev.\  {\bf 184} (1969) 1750.

\bibitem{Isham:1970gz}
  C.~J.~Isham, A.~Salam and J.~A.~Strathdee,
\emph{``Spontaneous breakdown of conformal symmetry,''}
  Phys.\ Lett.\ B {\bf 31} (1970) 300.

\bibitem{Fierz:1939ix}
  M.~Fierz, W.~Pauli,
\emph{``On relativistic wave equations for particles of arbitrary
spin in an electromagnetic field,''}
  Proc.\ Roy.\ Soc.\ Lond.\  {\bf A173 } (1939)  211-232.

\bibitem{Boulware:1973my}
  D.~G.~Boulware and S.~Deser,
\emph{``Can gravitation have a finite range?,''}
  Phys.\ Rev.\ D {\bf 6} (1972) 3368.




\bibitem{Boulware:1972zf}
  D.~G.~Boulware and S.~Deser,
\emph{``Inconsistency of finite range gravitation,''}
  Phys.\ Lett.\ B {\bf 40} (1972) 227.





\bibitem{deRham:2010kj}
  C.~de Rham, G.~Gabadadze and A.~J.~Tolley,
\emph{``Resummation of Massive Gravity,''}
  Phys.\ Rev.\ Lett.\  {\bf 106} (2011) 231101
  [arXiv:1011.1232 [hep-th]].

\bibitem{deRham:2011rn}
  C.~de Rham, G.~Gabadadze and A.~J.~Tolley,
\emph{``Ghost free Massive
 Gravity in the St\'uckelberg language,''}
  Phys.\ Lett.\ B {\bf 711} (2012) 190
  [arXiv:1107.3820 [hep-th]].



%



\bibitem{Hinterbichler:2012cn}
  K.~Hinterbichler and R.~A.~Rosen,
\emph{``Interacting Spin-2 Fields,''}
  JHEP {\bf 1207} (2012) 047
  [arXiv:1203.5783 [hep-th]].


\bibitem{Hassan:2012qv}
  S.~F.~Hassan, A.~Schmidt-May and M.~von Strauss,
\emph{``Proof of Consistency of Nonlinear Massive Gravity in the
St\'uckelberg Formulation,''}
  arXiv:1203.5283 [hep-th].








\bibitem{Hassan:2011hr}
  S.~F.~Hassan and R.~A.~Rosen,
\emph{``Resolving the
 Ghost Problem in non-Linear Massive Gravity,''}
  Phys.\ Rev.\ Lett.\  {\bf 108} (2012) 041101
  [arXiv:1106.3344 [hep-th]].


\bibitem{Hassan:2011vm}
  S.~F.~Hassan and R.~A.~Rosen,
\emph{``On Non-Linear Actions for Massive Gravity,''}
  JHEP {\bf 1107} (2011) 009
  [arXiv:1103.6055 [hep-th]].





\bibitem{Hassan:2011ea}
  S.~F.~Hassan and R.~A.~Rosen,
\emph{``Confirmation of the Secondary Constraint and Absence of
Ghost in Massive Gravity and Bimetric Gravity,''}
  JHEP {\bf 1204} (2012) 123
  [arXiv:1111.2070 [hep-th]].
%

\bibitem{Kluson:2012wf}
  J.~Kluson,
\emph{``Non-Linear Massive Gravity with Additional Primary
Constraint and Absence of Ghosts,''}
  Phys.\ Rev.\ D {\bf 86} (2012) 044024
  [arXiv:1204.2957 [hep-th]].


\bibitem{Hassan:2011tf}
  S.~F.~Hassan, R.~A.~Rosen and A.~Schmidt-May,
\emph{``Ghost-free Massive Gravity with a General Reference
Metric,''}
  JHEP {\bf 1202} (2012) 026
  [arXiv:1109.3230 [hep-th]].

\bibitem{Hassan:2011zd}
  S.~F.~Hassan and R.~A.~Rosen,
\emph{``Bimetric Gravity
 from Ghost-free Massive Gravity,''}
  JHEP {\bf 1202} (2012) 126
  [arXiv:1109.3515 [hep-th]].


\bibitem{Gruzinov:2011sq}
  A.~Gruzinov,
\emph{``All Fierz-Paulian massive gravity theories have ghosts or
superluminal modes,''}
  arXiv:1106.3972 [hep-th].



\bibitem{Burrage:2011cr}
  C.~Burrage, C.~de Rham, L.~Heisenberg and A.~J.~Tolley,
\emph{``Chronology Protection in Galileon Models and Massive
Gravity,''}
  JCAP {\bf 1207} (2012) 004
  [arXiv:1111.5549 [hep-th]].

\bibitem{deFromont:2013iwa}
  P.~de Fromont, C.~de Rham, L.~Heisenberg and A.~Matas,
\emph{``Superluminality in the Bi- and Multi- Galileon,''}
  arXiv:1303.0274 [hep-th].


\bibitem{Deser:2001dt}
  S.~Deser and A.~Waldron,
\emph{``Inconsistencies of massive charged gravitating higher
spins,''}
  Nucl.\ Phys.\ B {\bf 631} (2002) 369
  [hep-th/0112182].

\bibitem{Deser:2013uy}
  S.~Deser, M.~Sandora and A.~Waldron,
\emph{``Nonlinear Partially Massless from Massive Gravity?,''}
  Phys.\  Rev.\  D 87, {\bf 101501} (R) (2013)
  [arXiv:1301.5621 [hep-th]].


\bibitem{DeFelice:2012mx}
  A.~De Felice, A.~E.~Gumrukcuoglu and S.~Mukohyama,
\emph{``Massive gravity: nonlinear instability of the homogeneous
and isotropic universe,''}
  Phys.\ Rev.\ Lett.\  {\bf 109} (2012) 171101
  [arXiv:1206.2080 [hep-th]].

\bibitem{DeFelice:2013awa}
  A.~De Felice, A.~E.~Gümrükçüog(lu, C.~Lin and S.~Mukohyama,
\emph{``Nonlinear stability of cosmological solutions in massive
gravity,''}
  JCAP {\bf 1305} (2013) 035
  [arXiv:1303.4154 [hep-th]].

\bibitem{DeFelice:2013bxa}
  A.~De Felice, A.~E.~Gumrukcuoglu, C.~Lin and S.~Mukohyama,
\emph{``On the cosmology of massive gravity,''}
  arXiv:1304.0484 [hep-th].

\bibitem{DeFelice:2013tsa}
  A.~De Felice and S.~Mukohyama,
\emph{``Towards consistent extension of quasidilaton massive
gravity,''}
  arXiv:1306.5502 [hep-th].



\bibitem{Gourgoulhon:2007ue}
  E.~Gourgoulhon,
 \emph{``3+1 formalism and bases of numerical relativity,''}
  gr-qc/0703035 [GR-QC].


\bibitem{Arnowitt:1962hi}
  R.~L.~Arnowitt, S.~Deser, C.~W.~Misner,
 \emph{``The Dynamics of general
 relativity,''}
  [gr-qc/0405109].

\bibitem{Chaichian:2011sx}
  M.~Chaichian, M.~Oksanen and A.~Tureanu,
\emph{``Arnowitt-Deser-Misner representation and Hamiltonian
analysis of covariant renormalizable gravity,''}
  Eur.\ Phys.\ J.\ C {\bf 71} (2011) 1657
   [Erratum-ibid.\ C {\bf 71} (2011) 1736]
  [arXiv:1101.2843 [gr-qc]].


\bibitem{Peldan:1993hi}
  P.~Peldan,
\emph{``Actions for gravity,
 with generalizations: A Review,''}
  Class.\ Quant.\ Grav.\  {\bf 11} (1994) 1087
  [gr-qc/9305011].


\bibitem{Nicolai:1992xx}
  H.~Nicolai and H.~J.~Matschull,
\emph{``Aspects of canonical gravity and supergravity,''}
  J.\ Geom.\ Phys.\  {\bf 11} (1993) 15.


\bibitem{Henneaux:1983vi}
  M.~Henneaux,
\emph{``Poisson Brackets Of The Constraints In The Hamiltonian
Formulation Of Tetrad Gravity,''}
  Phys.\ Rev.\ D {\bf 27} (1983) 986.

\bibitem{Charap:1988vz}
  J.~M.~Charap, M.~Henneaux and J.~E.~Nelson,
\emph{``Explicit Form Of The Constraint Algebra In Tetrad
Gravity,''}
  Class.\ Quant.\ Grav.\  {\bf 5} (1988) 1405.

\bibitem{Yepez:2011bw}
  J.~Yepez,
\emph{``Einstein's vierbein field theory of curved space,''}
  arXiv:1106.2037 [gr-qc].


\bibitem{Damour:2002ws}
  T.~Damour and I.~I.~Kogan,
\emph{``Effective Lagrangians and universality classes of nonlinear
bigravity,''}
  Phys.\ Rev.\ D {\bf 66} (2002) 104024
  [hep-th/0206042].

\bibitem{Kluson:2013cy}
  J.~Kluson,
\emph{``Is Bimetric Gravity Really Ghost Free?,''}
  arXiv:1301.3296 [hep-th].



\bibitem{Kluson:2013lza}
  J.~Kluson,
\emph{``Hamiltonian Formalism of General Bimetric Gravity,''}
  arXiv:1303.1652 [hep-th].

\bibitem{Kluson:2012ps}
  J.~Kluson,
\emph{``Hamiltonian Formalism of Particular Bimetric Gravity
Model,''}
  arXiv:1211.6267 [hep-th].

\bibitem{Soloviev:2012wr}
  V.~O.~Soloviev and M.~V.~Tchichikina,
\emph{``Bigravity in Kuchar's Hamiltonian formalism. 1. The General
Case,''}
  arXiv:1211.6530 [hep-th].






\bibitem{Soloviev:2013mia}
  V.~O.~Soloviev and M.~V.~Tchichikina,
\emph{``Bigravity in Kuchar's Hamiltonian formalism. 2. The special
case,''}
  arXiv:1302.5096 [hep-th].









\bibitem{Henneaux:1992ig}
  M.~Henneaux and C.~Teitelboim,
\emph{``Quantization of gauge systems,''}
  Princeton, USA: Univ. Pr. (1992) 520 p


\bibitem{Comelli:2013txa}
  D.~Comelli, F.~Nesti and L.~Pilo,
\emph{``Massive gravity: a General Analysis,''}
  arXiv:1305.0236 [hep-th].

\bibitem{Comelli:2013paa}
  D.~Comelli, F.~Nesti and L.~Pilo,
\emph{``Weak Massive Gravity,''}
  arXiv:1302.4447 [hep-th].



\bibitem{Berezhiani:2013dw}
  L.~Berezhiani, G.~Chkareuli and G.~Gabadadze,
\emph{``Restricted Galileons,''}
  arXiv:1302.0549 [hep-th].

\bibitem{Berezhiani:2013dca}
  L.~Berezhiani, G.~Chkareuli, C.~de Rham, G.~Gabadadze and A.~J.~Tolley,
\emph{``Mixed Galileons and Spherically Symmetric Solutions,''}
  arXiv:1305.0271 [hep-th].

\bibitem{Chamseddine:1978yu}
  A.~H.~Chamseddine, A.~Salam and J.~A.~Strathdee,
\emph{``Strong Gravity and Supersymmetry,''}
  Nucl.\ Phys.\ B {\bf 136} (1978) 248.

\bibitem{Chamseddine:2004dh}
  A.~H.~Chamseddine,
\emph{``Matrix gravity and massive colored gravitons,''}
  Phys.\ Rev.\ D {\bf 70} (2004) 084006
  [hep-th/0406263].

\bibitem{Chamseddine:2003ft}
  A.~H.~Chamseddine,
\emph{``Spontaneous symmetry breaking for massive spin two
interacting with gravity,''}
  Phys.\ Lett.\ B {\bf 557} (2003) 247
  [hep-th/0301014].

\bibitem{Alexandrov:2013rxa}
  S.~Alexandrov,
\emph{``Canonical structure of Tetrad Bimetric Gravity,''}
  Gen.\ Rel.\ Grav.\  {\bf 46} (2014) 1639
  [arXiv:1308.6586 [hep-th]].


\bibitem{DeFelice:2014nja}
  A.~De Felice, A.~E.~Gumrukcuoglu, S.~Mukohyama, N.~Tanahashi and T.~Tanaka,
\emph{``Viable cosmology in bimetric theory,''}
  arXiv:1404.0008 [hep-th].





\end{thebibliography}
\end{document}